# Observation of Quantum Effects in sub Kelvin Cold Reactions

A.B. Henson, S. Gersten, Y. Shagam, J. Narevicius, E. Narevicius[*]

Department of Chemical Physics, Weizmann Institute of Science, Rehovot, Israel

[*]To whom correspondence should be addressed E-mail: edn@weizmann.ac.il

**Abstract**

There has been a long-standing quest to observe chemical reactions at low temperatures where reaction rates and pathways are governed by quantum mechanical effects. So far this field of Quantum Chemistry has been dominated by theory. The difficulty has been to realize in the laboratory low enough collisional velocities between neutral reactants, so that the quantum wave nature could be observed. We report here the first realization of merged neutral supersonic beams, and the observation of clear quantum effects in the resulting reactions. We observe orbiting resonances in the Penning ionization reaction of argon and molecular hydrogen with metastable helium leading to a sharp increase in the absolute reaction rate in the energy range corresponding to a few degrees kelvin down to 10 mK. Our method is widely applicable to many canonical chemical reactions, and will enable a breakthrough in the experimental study of Quantum Chemistry.

One of the fundamental traits of Quantum Chemistry is the effect of wave-like phenomena in reactions between atoms and molecules. So far studies in this field were mainly confined to theory, due to the experimental challenge of achieving low enough collision energies where reactants de Broglie wavelength approaches the characteristic interaction length scale. Among the experiments that reported observations of reactions at temperatures of 1 K or below (*1-4*) the evidence of quantum effects in the reaction of ultracold KRb molecules was shown by Ospelkaus et al. (*3*). It is anticipated that in the cold regime quantum effects such as tunneling through reaction or centrifugal barriers will dominate the reaction dynamics (*5*). One of the most striking manifestations of quantum tunneling is the formation of scattering resonance states that can dramatically affect the reaction rate. Quasibound resonance states arise in the continuum part of the spectrum, making them accessible to scattering experiments where the collision energy can be tuned. Detection of scattering resonances in reactions will allow an accurate determination of the interaction potentials governing the collision dynamics. Such measurements will constrain various theoretical models and methods, such as electronic structure calculations (that due to complexity always involve approximations), and molecular dynamics simulations that necessarily require quantum treatment at these low temperatures. Note, that dynamical Feshbach resonances have been measured in the differential cross section of the F+$H_2$ reaction (*6*) yet so far tunneling resonances have not been observed in low temperature reactive scattering.

We report here the first observation of scattering resonances in the Penning ionization reaction of argon and molecular hydrogen with metastable helium atoms. We resolve peaks in the energy dependent reaction rate arising from the formation of metastable collision complexes via quantum tunneling through an angular momentum barrier. The experimental breakthrough that allowed us the observation of tunneling resonances is in our ability to tune collision temperatures down to 10 mK, despite employing the conventional supersonic molecular and atomic beams without any additional cooling stage. At 10 mK collision temperature the corresponding de Broglie wavelength in the case of molecular hydrogen Penning ionization is 38 nm. The temperatures we achieve are more than three orders of magnitude lower than those attainable in state-of-the-art systems featuring collision energy tunability. In order to understand such a dramatic effect one needs to realize that the relative velocity, or collision energy, can be reduced by merging two beams. The advantage of this approach was recently theoretically investigated by Wei et al (*7*) and we realized this idea by bending the trajectories in one of the beams in a curved magnetic quadrupole guide.

The history of angular momentum barrier tunneling, or equivalently called orbiting resonances, dates far back to the origins of collisional theory. Boltzmann postulated that a collisional metastable complex forms during recombination reactions via an interaction with a third body (*8*). This possibility was later ruled out as improbable. Bunker recognized that the metastable intermediate state, formed by the colliding atoms and molecules, can be trapped by the potential barrier which emerges from the contribution of the centrifugal potential (*9*). In the case of elastic scattering the effect of the orbiting resonances on the elastic collision cross section has been experimentally measured by making a very clever choice of species leading to a system with a low reduced mass and a weak van der Waals interaction strength (*10-12*). At these favorable conditions orbiting resonances are formed at high collision energies equivalent to about 10 K.

Orbiting resonances have been predicted to occur in many reactions (*13-17*). Indeed, the appearance of resonances in low energy collisions is a rule rather than an exception. However the conditions necessary for the observation of scattering resonances are confounded by the lowest collision energies available in experiments, as well as the requirement for a resolution high enough to resolve individual resonance states. Practically, one needs to scan reaction rates in an energy range corresponding to temperatures from a few degrees kelvin to a few millikelvin. Below several mK only the s-wave scattering channel remains open and the Wigner threshold law (*18*) settles in with the monotonic behavior of reaction rate re-established.

Among the many available methods that permit tuning of the collision energy none have reached temperatures sufficiently low to make the observation of resonances, in the cold regime, routinely possible. One way to obtain cold colliding species is by using a cryogenic cell (*19, 20*) or within a uniform supersonic expansion with carefully controlled temperature and density (*21, 22*). Another school of thought, motivated by the pioneering experiments conducted by Herschbach and Y. T. Lee (*23*) involving the use of two crossed supersonic beams, aims to control the collision energy by varying the angle of collision or the velocity of one of the beams. Recently adjusting the crossing angle has been used to measure a reaction cross section down to a temperature of 20 K (*24*). Meijer and colleagues used pulsed electric field decelerated supersonic beams to study inelastic scattering (*25*). Inelastic scattering experiments were also performed with magnetically trapped atoms and molecules (*26, 27*) with the lowest temperature reported at 5 K (*28*).

In our approach we were able to break through the 1 K collision temperature barrier by merging two fast supersonic beams. As a result the relative velocity vanishes in a moving frame of reference when beams of equal velocities merge. Similar methods have been used in ion chemistry for almost two decades (*29*). In contrast to charged particles where trajectories can be easily controlled via interaction with modest electric fields, the neutral particles center of mass motion is far more challenging to control. We utilize the Zeeman effect on a paramagnetic species in order to direct them through a curved magnetic quadrupole guide. At the exit of the quadrupole the paramagnetic particle beam merges with another supersonic beam that has travelled in a straight line from a separate supersonic source. Importantly, there is no restriction on the nature of the second beam and any atom or molecule, which can be entrained into the supersonic beam, can be used. By varying the relative velocities between the two beams we continuously tune collisional energies from 700 K down to 10 mK.

In our study of low energy Penning ionization reactions we merge a supersonic beam of metastable helium with a supersonic beam containing either argon or molecular hydrogen. The excited $^3$S state of helium is located 19.8 eV above the ground state. When a metastable helium atom collides with another atom or molecule with an ionization energy lower than 19.8 eV, a charge transfer process takes place where an electron from the neutral species "jumps" into the vacancy of the 1s orbital of helium coinciding with the ejection of an electron from the 2s orbital to the continuum.

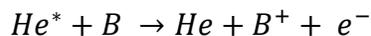
$$He^* + B \rightarrow He + B^+ + e^-$$

Penning ionization has been studied in detail at higher energies with many interesting results summarized in a review article by Siska (*30*). Penning ionization reactions have different entrance and exit channels (Fig. 1). In the entrance channel the interaction between metastable helium and another atom or molecule can be described using a complex optical potential. The real part of the interatomic potential contains the appropriate long range van der Waals interaction whose leading term scales as R$^{-6}$, where R is the distance between metastable He atom and the colliding atom or molecule. At short distances the repulsion term takes over whereas at intermediate distances there is a shallow van der Waals well. The complex part of the potential $\Gamma(R)/2$ is related to the ionization probability at a given internuclear distance. Since the charge transfer probability decays rapidly with separation it is usually modeled by a single exponential term. Electronic motion is much faster compared to the nuclear motion and the autoionization process can be viewed as a vertical process within the Born Oppenheimer approximation. The resulting ion and neutral helium interaction is described by a neutral-ion potential surface in the exit channel.

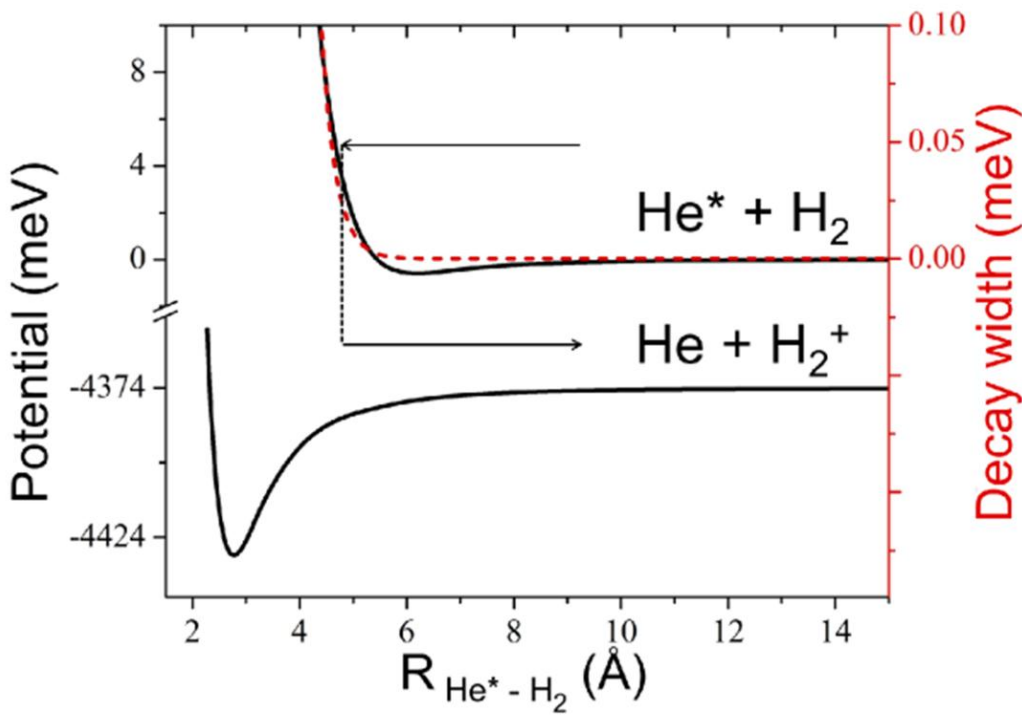

**Fig. 1.** The spherical part of the interaction potentials involved in the Penning ionization reaction of metastable He ($^3$S) and molecular hydrogen. The entrance channel includes both the real part of the potential in the upper solid black line as well as autoionization width in the dashed red line. The lower black curve represents the ion-neutral exit channel interaction potential.

A schematic of the experimental system is shown in Fig. 2A. The pulsed metastable helium supersonic beam is created with an Even-Lavie valve (*31*), cooled down to 55 K. Immediately after the valve there is a Dielectric Barrier Discharge (DBD) which is used to excite the ground state helium to the $2^3$S level. The beam has a mean velocity of ~770 m/s with a standard deviation of 15 m/s corresponding to a temperature of 100 mK in the moving frame of reference. The beam then passes through a 4mm diameter skimmer located 10 cm from our valve orifice and enters a 20 cm long magnetic quadrupole which has a 10$^o$ curve with a curvature radius of 114.7 cm. We create a quadrupole magnetic field by passing a current pulse through 1mm diameter copper wires arranged in quadratures, as shown in Fig. 2B, each quadrature consists of 9 wires in a 3x3 pattern. At the peak current of 1000A the transverse quadrupole trap depth is about 2.7 K and 3x3 mm$^2$ in size. Only low field seeking Zeeman sublevels are confined in two dimensions during the transit through the quadrupole guide. As such, the metastable helium beam that leaves the magnetic quadrupole is 100% spin polarized in a single quantum state with the projection of the total angular momentum on the quantization axis, $m_J$ = 1, originating from the $^3$S electronic state manifold. At the end of the quadrupole the metastable beam is merged with a second supersonic beam, also created with a temperature controlled Even-Lavie valve, and collimated with a 4mm skimmer. This second beam contained molecular hydrogen in one experiment and atomic argon in another, either as neat gasses or seeded in a noble carrier gas. By changing the composition of the gas mixtures as well as the temperature of the valve (between 355 K and 120 K) we changed the relative mean velocity between the beams from over 1000 m/s down to zero. At zero relative velocity between the beams the residual collision energy stems from the finite velocity distribution of the supersonic beams at the collision volume.

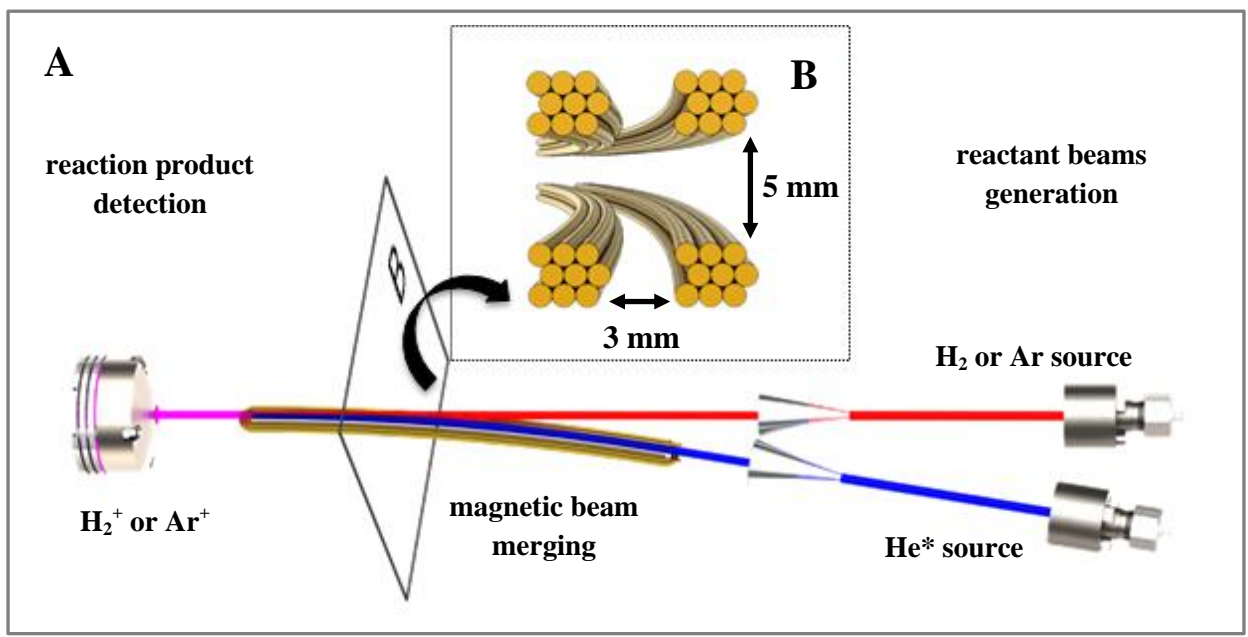

Fig. 2. (A) Schematic of the experimental system without vacuum chambers, showing the two source supersonic valves followed by two skimmers, the curved magnetic quadrupole guide with its assembly and the quadrupole mass spectrometer (QMS) entrance at the end. The blue beam depicts the magnetically guided beam whereas the red beam is unaffected, the merged volume is in purple. Also shown (B) is a front view of the quadrupole guide.

We measure the time-of-flight signal for both reactant beams as well as the product. The metastable beam is measured using a Micro Channel Plate (MCP), whereas the ground state beam is measured using an ionizing quadrupole mass spectrometer (QMS). To measure the product ion we turn off the ionizing element of the QMS in order to observe the ions formed in the chemi-ionization collisions. We first divide the ion signal by the product of the area of both neutral beams and then normalize using the data from earlier high collision energy experiments (*32, 33*). Thus we are able to present our results for hydrogen (Fig. 3A) and for argon (Fig. 3B) on the absolute reaction rate scale.

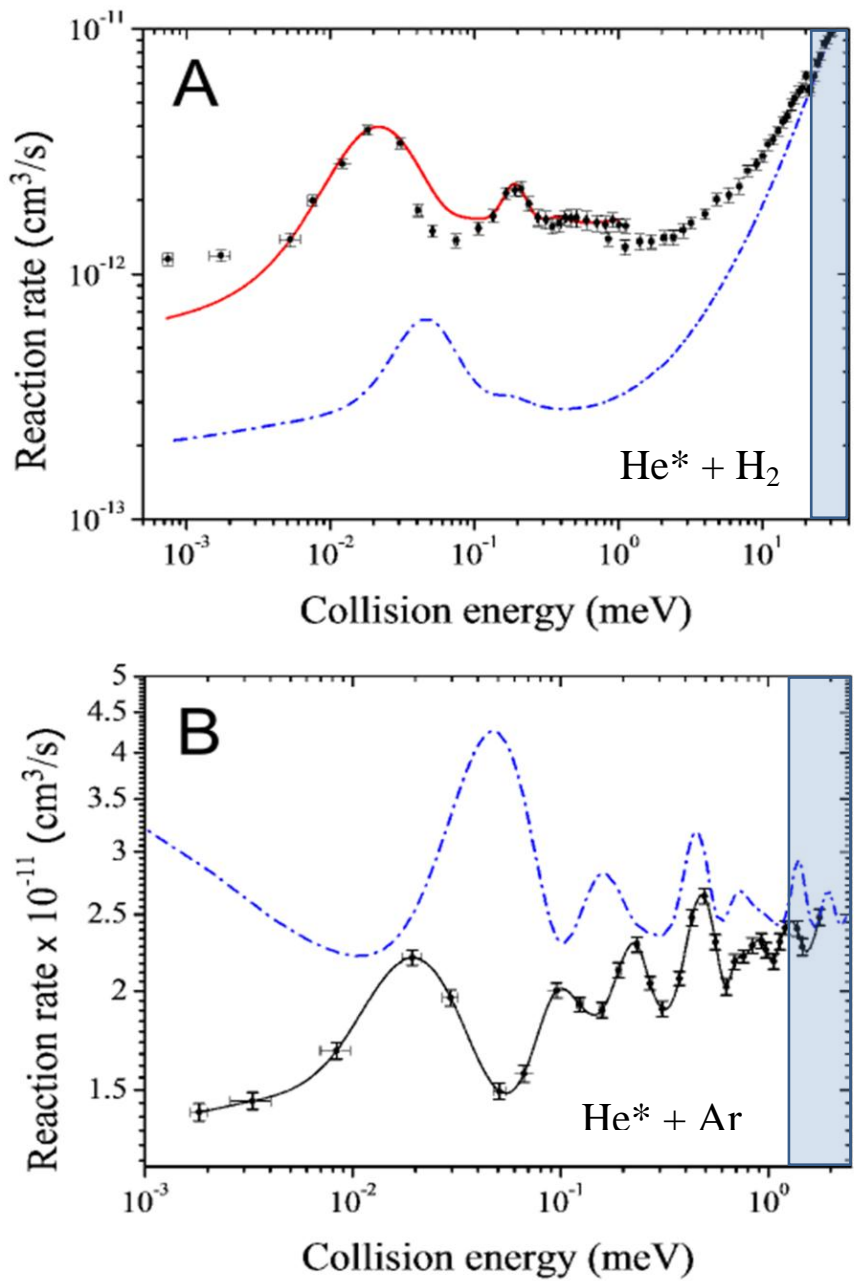

**Fig. 3.** (A) Reaction rate measurements for the ($^3$S) He$^*$ and H$_2$ Penning ionization reaction are shown in black with error bars. Blue dash-dot line is the reaction rate calculated using the latest "experimental" potential from Ref (33). Red solid line is the calculated reaction rate using the Tang-Toennies potential with parameters that give the best fit to our measured results. A second slower decay term is included in the imaginary potential part in order to adjust the reaction rate value at low collision energies. (B) Reaction rate measurements for the ($^3$S) He$^*$ and Ar Penning ionization reaction are shown in black with error bars, and the solid black line is a guide to the eye. The blue dash-dot line is the reaction rate calculated using the latest "experimental" potential from Ref (32). Shaded areas in both figures correspond to the energy range where our results overlap with the earlier measurements.

At higher collision energies, above 1 meV (23 K) and 20 meV (460 K) for Ar and $H_2$ systems respectively, our results are in very good agreement with earlier experimental measurements. A classical treatment of the collisional process is sufficient to explain the main results above this energy. The reaction rate falls at lower velocities since the inner classical turning point position scales with energy. For lower energies the classical turning is positioned at larger internuclear separation where the ionization rate is lower. At about 5 meV and 1 meV, in the cases of Ar and $H_2$ ionization respectively, the reaction rate levels off since the decrease of the ionization rate is compensated by longer collision times.

At energies below 1 meV collisions enter the quantum regime and we observe a series of peaks in the total reaction rate as a function of energy. Peaks appear both in the case of Penning ionization of argon and of molecular hydrogen. The number of peaks and the peak spacing scales with the reduced mass indicating that the quantization of the internuclear coordinate is involved. To elucidate the origins of the resonant structures we have performed numerical simulations of the collision process. We numerically integrate the Schrodinger equation with a complex potential and for each partial wave determine the complex phase shift from the asymptotic behavior of the numerical solution. The total ionization reaction rate is given by a weighted sum over the contribution of individual partial waves.

We used the latest potential energy surfaces constructed using both ab-initio calculation results and experimental data obtained at higher energies. As one can see from Fig. 3 the old potential correctly captures the higher energy behavior, yet fails to reproduce the positions of the resonance states. Since the interaction potentials of metastable helium with molecular hydrogen have been thoroughly investigated using ab-initio methods we chose this system to try and fit the potential surface parameters so as to match the observed resonance positions. We used a Tang-Toennies potential (*34*), starting with parameters derived from the "experimental" potential surface. We were able to match the resonance positions only after we increased the van der Waals potential well depth by a factor of 1.8 as well as move the well minimum position by 1Å. Such sensitivity to the parameters of the potential surface illustrates the importance of measuring the resonance states in order to obtain the correct long range interaction behavior. Furthermore it will stimulate theoretical investigations of the metastable helium molecular hydrogen potential due to the system's simplicity as it involves only four electrons and thus can serve as a benchmark for different ab-initio methods. In addition to having to correct the positions of the resonance peaks there was also a discrepancy, at low energies, of almost an order of magnitude between the measured and calculated reaction rates. The observed higher reaction rate can be reproduced in our calculations by the addition of a slower decay term to the imaginary part of the potential. This raises the question whether the estimation of an autoionization width as a single decaying exponent is sufficiently accurate. This discrepancy could possibly arise from the underestimation of atomic and molecular orbital overlap at large separations.

We have also calculated the resonance wave functions using discrete variable representation (DVR) (*35*) and complex scaling methods (*36*). In the case of molecular hydrogen Penning ionization the resonant states appear in the partial waves associated with the total angular momentum l =5 and 6 (Fig. 4). The orbiting resonances wave functions are highly delocalized, extending to more than 10 Å. Interestingly, the orbiting resonance associated with the l=6 partial wave channel is located above the energy of the centrifugal barrier. Such a resonance is formed by a quantum reflection off, and not tunneling through, the angular momentum barrier.

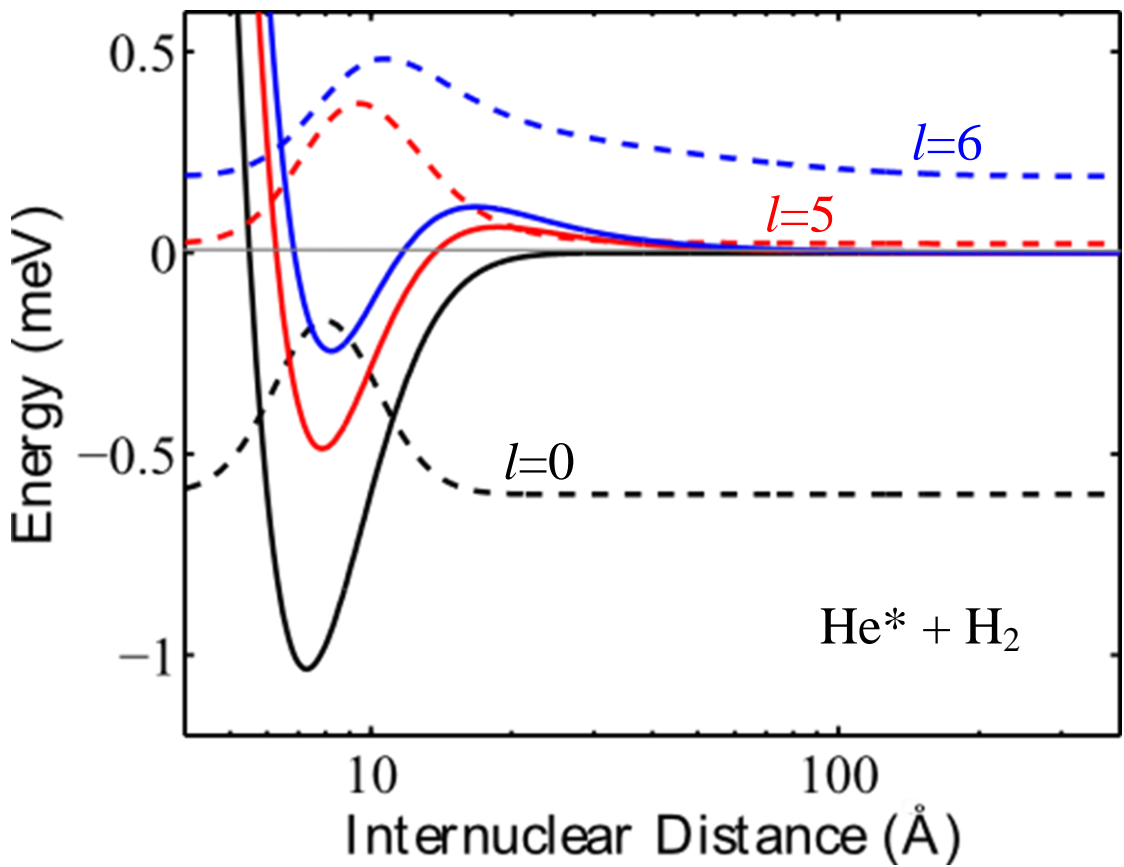

**Fig. 4.** Potentials and wave-functions for the entrance channel of ($^3$S) He$^*$ and H$_2$ reaction. Potentials for different contributing angular momentum terms are shown in solid lines with the corresponding absolute value of wave-functions in dashed lines. The black bottom-most lines are for the potential with l=0. Lines above that in red are for l=5 and the top-most blue lines are for the l=6 partial wave.

The resonance wave functions illustrate the mechanism of low energy resonant chemi-ionization. Classically, one would expect that collisions of particles at energies below the centrifugal barrier should not contribute to the ionization process. However, at collision energies that match positions of the resonance states, particles tunnel through the centrifugal barrier and become trapped. In such a case the probability to find colliding particles, described by the absolute value of corresponding wave function, is enhanced inside of the centrifugal barrier within the region where the autoionization rate is higher.

Our results show the observation of a clear quantum effect in reactions below 1K. We have observed orbiting resonances in the Penning ionization reaction of molecular hydrogen and argon with metastable helium that are formed by tunneling through an angular momentum potential barrier. As a result we observe sharp peaks in the reaction rate measurements as a function of collision energy. The observation of quantum effects was enabled by our method that is based on the magnetic merging of two fast supersonic beams. We take advantage of having the cold collisions take place within the moving frame of reference and extend the experimentally attainable collision energy by three orders of magnitude down to the temperatures of 10 mK. Our method is general for any gas phase reactions that have one paramagnetic reagent. This will allow the implementation of our method to species of vast scientific interest in the fields of astrochemistry and cold collisions with species such as H, O, N, F, OH, $O_2$ and opens the door to the highly anticipated field of cold chemistry. We plan to use our method to study the canonical $F+H_2$ reaction in low energy regime.

37   We acknowledge Uzi Even for many helpful discussions and sound advice.  We thank Roee Ozeri for most fruitful discussions. We thank Mark Raizen, Ron Naaman, Yehiam Prior and Barak Dayan for a careful reading of the manuscript. This research was made possible, in part, by the historic generosity of the Harold Perlman Family. E.N. acknowledges support from the Israel Science Foundation and Minerva foundation. We thank Shlomo Assayag and Michael Vinetsky from the Weizmann machine shop for assistance in designing and manufacturing of our vacuum chamber.